\documentclass[aps,prl,twocolumn,groupedaddress]{revtex4-1}  
\usepackage{graphicx}  
\usepackage{dcolumn}   
\usepackage{bm}        
\usepackage{amssymb,amsmath}   
\usepackage{verbatim}

\usepackage{amsmath,amssymb,amsfonts}

\usepackage{float} 
\usepackage{cleveref}              

\usepackage[usenames,dvipsnames]{xcolor}
\usepackage{epsfig}
\usepackage{epstopdf}
\usepackage{dcolumn}
\usepackage{tikz}
\usetikzlibrary{shapes.geometric, arrows}
\usepackage{upgreek}
\usepackage{setspace}
\usepackage{enumitem}
\usepackage{array,multirow,bigdelim}

\def\be{\begin{equation}}
\def\ee{\end{equation}}
\def\ba{\begin{eqnarray}}
\def\ea{\end{eqnarray}}

\def\a{\alpha}

\def\CP1{\mathbb{CP}^1}
\def\SL2C{\mathrm{SL}(2,\mathbb{C})}

\def\Z2{\mathbb{Z}_2}

\def\su2{{SU(2)}}

\def\a{{\alpha}}

\def\[{\left[}
\def\]{\right]}

\def\L{\Lambda}

\def\s{\sigma}
\def\a{\alpha}

\def\({\left(}
\def\){\right)}
\def\[{\left[}
\def\]{\right]}

\def\<{\langle}
\def\>{\rangle}

\def\i2{\frac{i}{2}}

\def\2F1{\,_2{\rm F}_1}

\newcommand{\beq}{\begin{equation}}
\newcommand{\eeq}{\end{equation}}
\newcommand{\beqq}{\begin{equation*}}
\newcommand{\eeqq}{\end{equation*}}
\newcommand\beqa{\begin{eqnarray}}
\newcommand\eeqa{\end{eqnarray}}
\newcommand\beqaa{\begin{eqnarray*}}
\newcommand\eeqaa{\end{eqnarray*}}
\newcommand\bea{\begin{array}}
\newcommand\eea{\end{array}}

\begin{document}

\widetext


\title{New  Factorization Relations for Non-Linear Sigma Model Amplitudes}
\author{N. E. J. Bjerrum-Bohr$^1$, Humberto Gomez$^{1,2}$, Andreas Helset$^1$}
\affiliation{\smallskip$^1$Niels Bohr International Academy and Discovery Center\\ Niels Bohr Insitute, University of Copenhagen\\
Blegdamsvej 17, DK-2100 Copenhagen \O, Denmark\smallskip \\ 
$^2$Facultad\! de\! Ciencias$,$\! Basicas\! Universidad\!\\ Santiago\! de\! Cali,\!
Calle 5 $N^\circ$\!\! 62-00 Barrio Pampalinda\\ Cali, Valle, Colombia}

\begin{abstract}
We obtain novel factorization identities for non-linear sigma model amplitudes using a new integrand in the CHY double-cover prescription. We find that it is possible to write very compact relations using only longitudinal degrees of freedom. We discuss implications for on-shell recursion.
\end{abstract}

\maketitle
\section{Introduction}\vskip-0.1cm
Cachazo, He and Yuan invented in ref. \cite{Cachazo:2013gna} a new method for calculating S-matrix elements. This formalism has numerous applications and many interesting connections, see for instance~refs. \cite{
Mason:2013sva,Berkovits:2013xba,Bjerrum-Bohr:2014qwa}. The CHY construction was formally proven by Dolan and Goddard in ref. \cite{Dolan:2013isa}.\\[5pt]
The main ingredients are the $n$-point scattering equations
\noindent\vskip-0.5cm
\begin{align}
	0=S_a\equiv\hspace{-0.2cm}\sum_{b=1,b\neq a}^n \hspace{-0.2cm}\frac{s_{ab}}{z_{ab}},\quad z_{ab}\equiv z_a - z_b, \,\,\,\, s_{ab}\equiv2k_a \cdot k_b,
\end{align}\vskip-0.2cm
\noindent
where $z_{a}$ are auxiliary variables on the Riemann sphere and $k_a$ are momenta. In the CHY formalism one has to integrate over a contour containing the $(n-3)!$ independent solutions  of the scattering equations. \\[5pt]
As computations in the CHY formalism grow factorially in complexity with $n$, integration rules have been developed at tree \cite{Baadsgaard:2015voa,Cachazo:2015nwa,Bjerrum-Bohr:2016juj,Cross-ratio} and loop level \cite{Baadsgaard:2015hia}, so that analytical results for amplitudes can be derived without solving the scattering equations explicitly.\\[5pt]
Recently, the CHY formalism was reformulated by one of us in the context of a double cover \cite{Gomez:2016bmv} (called the 
'$\Lambda$-formalism' in refs. \cite{Gomez:2016bmv,Cardona:2016bpi}).
Here, the basic variables are elements of $\mathbb{CP}^2$, and not $\mathbb{CP}^1$ as 
in the original CHY formalism.
One advantage of the extra machinery is that amplitudes in the double-cover formulation naturally
factorize into smaller $\mathbb{CP}^1$ pieces, and this is a useful laboratory for deriving new amplitude identities. \\[5pt]
We will start by reviewing the CHY formalism for the non-linear sigma model (NLSM) and provide an alternative formulation that employs a new integrand. Next, we will show how the double-cover formalism naturally factorizes this new CHY formulation in a surprising way. \\[-19pt]
\section{A new CHY integrand}
\vskip-0.1cmAs explained in ref. \cite{Cachazo:2014xea}, the flavor-ordered partial $U(N)$ non-linear sigma model amplitude in the scattering equation framework is given by 
the contour integral\\[-10pt]
\begin{align}
    A_n(\a)&=\int d\mu_n H_n(\a)\,,
\\[-5pt] \nonumber 
d\mu_n&\equiv   (z_{ij} \, z_{jk}\, z_{ki})^2
\prod_{a=1 \atop a \neq \{i,j,k\}}^n \frac{dz_a}{S_a},
\end{align}\\[-10pt]
\noindent 
where $(\a)=(\a(1),...,\a(n))$ denotes a partial ordering. 
The integrand is given by
\noindent
\begin{align}\hskip-1.9cm
H_n (\a)  = 
{\rm  PT}{(\a)}
\times ({\rm Pf}^\prime \mathsf{A})^2, 
\end{align}\\[-35pt]
\begin{align}
{\rm PT}(\a) &\equiv  \frac{1}{z_{\a(1) \a(2)} z_{\a(2) \a(3)}  \cdots z_{\a(n) \a(1)}}, \\
\label{eq:defredpfaffian}
{\rm Pf}^\prime \mathsf{A} &\equiv \frac{(-1)^{i+j}}{ z_{ij} }\,\,{\rm Pf}[(\mathsf{A})^{ij}_{ij}].
\end{align}\\[-10pt]
Here ${\rm PT}(\a) $ and ${\rm Pf}^\prime \mathsf{A} $ are the Parke-Taylor $(\rm PT)$ factor and the reduced Pfaffian of matrix $\mathsf{A}$, respectively.
The $n\times n$ anti-symmetric matrix, $\mathsf{A}$, is defined as, \\[-10pt]
\begin{align}\displaystyle
\mathsf{A}_{ab}\equiv \displaystyle \frac{s_{ab}}{z_{ab}} \quad &{\rm for} \,\, a\neq b,\ \ \ {\rm and}\ \ \ 
        \mathsf{A}_{ab}\equiv0 &{\rm for} \,\, a=b.
\end{align}\\[-10pt]
In general, $(\mathsf{A})^{i_1\cdots i_p}_{j_1\cdots j_p}$ denotes the reduced matrix obtained by removing the rows, $\{ i_1,...,i_p \}$, and columns, $\{ j_1,...,j_p \}$, from $\mathsf{A}$.
Note that when the number of external particles $n$ is odd, ${\rm Pf}^\prime \mathsf{A} =0$, and $A_n(\a)$ vanishes.\\[5pt]
Using \cref{eq:defredpfaffian}, we have 
\begin{align}
({\rm Pf}^\prime \mathsf{A})^2 = \frac{(-1)^{i+j+m+p}}{z_{ij} \, z_{mp}} \, {\rm Pf}[(\mathsf{A})^{ij}_{ij}]\times  {\rm Pf}[(\mathsf{A})^{mp}_{mp}].
\end{align}\\[-10pt]
With the choice $\{i,j\} = \{m,p\}$, this product of Pfaffians becomes a determinant,\\[-10pt]
\begin{align}
({\rm Pf}^\prime \mathsf{A})^2 = -{\rm PT}(m,p)\, {\rm det} [(\mathsf{A})^{mp}_{mp}]. 
\end{align}\\[-12pt]
We will now discuss the following new matrix identities.
On the support of the scattering equations and the massless condition, $\{S_a=0, k_a^2=0\}$, we find when $m \neq p \neq q$
\begin{align}
	{\rm Pf} [(\mathsf{A})^{mp}_{mp} ] \times {\rm Pf} [(\mathsf{A})^{pq}_{pq}] &=  {\rm det} [(\mathsf{A})^{mp}_{pq}], \\
{\rm det} [(\mathsf{A})^{mp}_{pq}] &= 0  \quad {\rm if } \, n \,\,{\rm odd} .
\end{align}
\noindent
 A proof of these identities will be provided in ref. \cite{preparation}.
Using the non-antisymmetric matrix, $(\mathsf{A})^{ij}_{jk}$, we define the objects (with $i<j<k$)\\[-16pt]
\begin{align}
 A_n' (\a) 
 &=\int d\mu_n\,\,
{\rm  PT}{(\a)} \, \frac{(-1)^{i+k} } {z_{ij}\, z_{jk}} \, 
{\rm det} [(\mathsf{A})^{ij}_{jk}], ~~ \label{newAmp1}\\
 A_n^{(ij)}(\a)
 &=\int d\mu_n\,\,
{\rm  PT}{(\a)} \, \frac{(-1)^{i+j} } {z_{ij}} \, 
{\rm det} [(\mathsf{A})^{i}_{j}] . ~~ \label{newAmp2}
\end{align}\\[-10pt]
\noindent
Note that in \cref{newAmp1,newAmp2} we have reduced the  
$\mathsf{A}$ matrix with the indices $\{i,j,k\}$ associated with the Faddeev-Popov determinant. 
This gauge choice will be convenient later.
We now have the following equality,
\begin{align}
     A_n'(\a)=A_n(\a),
\end{align}
\noindent
when all particles are on-shell. When there are off-shell particles, the identity is true only if the number of particles is even. 
When the number of particles is odd and there are off-shell particles, one has $A_n(\a)=0$ while $A_n'(\a)\neq0$. Since the $\mathsf{A}$ matrix has co-rank $2$ on the support of the scattering equations and the massless condition, $\{ S_a=0,k_a^2=0\}$, $ A_n^{(ij)}(\a)$ vanishes trivially. However, when there are off-shell particles the amplitude $ A_n^{(ij)}(\a)$ is no longer zero.\\[5pt]
These observations will be crucial in obtaining the new factorization relations.\\[-25pt]
\section{The double-cover representation}
In the  double-cover version of the CHY construction, the $n$-point amplitude is 
given as a contour integral on the double-covered Riemann sphere with $n$ punctures.
The pairs $(\sigma_1,y_1), (\sigma_2,y_2),\ldots, (\sigma_n,y_n)$
provide the new set of doubled variables restricted to the curves\\[-12pt]
\noindent
\begin{align}
	0 = C_a \equiv y_a^2 - \sigma_a^2 + \Lambda^2 \qquad \textrm{for } a = 1,\ldots,n.
\end{align}\\[-12pt]
A translation table has been worked out in detail in ref. \cite{Gomez:2016bmv}. The double-cover formulation of the NLSM is given by the integral\\[-13pt]
\begin{align}\label{YMgeneric}
\hspace{-0.2cm}
A_n(\a) &= \int_\Gamma d\mu^\L_n \, 
 \frac{(-1)\,\Delta{(ijk)} \, \Delta{(ijk|r)}} {S^\tau_r } \times
{\cal I}_n(\a) ,~~ \nonumber\\[-3pt]
d\mu_n^{\!\,\L}&\equiv
\frac{1}{2^2}
\frac{d\L}{\L} \, \prod_{a=1}^n \frac{y_a dy_a}{C_a} \!\times\! \!\!\!\!
\prod^n_{d=1 \atop d\neq i,j,k,r}\!\!\frac{d\s_{\!d}}{ S^{\tau}_d},
\end{align}\\[-15pt]
 \\[-30pt]
\begin{align}
\!\!\!\tau{(a,b)}\!&\equiv \frac{1}{2\,\s_{ab}}\left( \frac{y_a + y_b+ \s_{ab}}{y_a}\right),\ \
S_a^{\tau}\equiv 
\sum_{b=1 \atop b\neq a}^n s_{ab} \,\tau{(a,b)}\,, ~~\nonumber \\[-5pt]
\Delta_{\rm }(ijk)&\equiv  \big( \tau(i,j)\,\tau(j,k)\,\tau(k,i)\big)^{-1} \, ,~~ 
\\
\Delta_{\rm }(ijk|r)&\equiv 
\s_i\Delta_{\rm }(jkr)
- \s_r \Delta_{\rm }(ijk) +\s_k  \Delta_{\rm }(rij) - \s_j  \Delta_{\rm }(kri). ~~\nonumber
\end{align} 
The  $\Gamma$ contour is defined by the $2n-3$ equations
\begin{align}
\hspace{-0.4cm}
\L&=0, \ \ \ \ \  S^{\tau}_d(\s,y)=0, \ \ \ \ \ C_a=0,
\end{align}
for $d\neq \{i,j,k,r\}\,$ and  $a =1,...,n.$\\[5pt]
The integrand is given by 
\begin{align}
\!\!\!\!{\cal I}_n(\a) \!=\! - \!{\rm PT^\tau} (\a)\!  \prod_{a=1}^n \!  \frac{(y\s)_a}{y_a}  \, {\rm PT}^T(m,p) \, {\rm det} [(\mathsf{A}^\L)^{mp}_{mp}],\!\!\!\!\!
\end{align}
\noindent
where $(y\s)_a\equiv y_a+\s_a$. 
To obtain the kinematic matrix and the Parke-Taylor factors we need to do the following replacements
\begin{align}
	\mathsf{A} &\to \mathsf{A}^\L ,\ {\rm and}\  {\rm PT} \to {\rm PT}^T \ \ {\rm for \  \ } z_{ab} \rightarrow T_{ab}^{-1} , \\
	 {\rm PT} &\to  {\rm PT}^\tau\quad \hskip2.35cm{\rm for \ \ }z_{ab}  \rightarrow \tau(a,b)^{-1} ,
\end{align}
with $T_{ab}\equiv \frac{1}{(y\s)_a-(y\s)_b }$.
Analogous to \cref{newAmp1}, we can now write down a new form for the integrand
\begin{align}
\!\!\!\!\!\!{\cal I}_n'(\a)\! =\!  {\rm PT^\tau} (\a) \prod_{a=1}^n   \!\!\frac{(y\s)_a}{y_a}  (-1)^{i+k} T_{ij}T_{jk} \, {\rm det} [(\mathsf{A}^\L)^{ij}_{jk}],\!\!\!\!\!
\end{align}\\[-10pt]
where $\{i,j,k\}$ are the same labels as in $\Delta_{\rm }(ijk)\, \Delta_{\rm }(ijk|r)$.
For more details on the double-cover prescription, see refs. \cite{Gomez:2016bmv,Gomez:2018cqg,preparation}.\\[-36pt]
\section{Factorization}\label{Factorization}
Let us start by considering the four-point amplitude, $A_4'(1,2,3,4)$, with the gauge fixing $(ijk|r)=(123|4)$. We will denote sums of cyclically-consecutive external momenta (modulo the total number of external momenta) by $P_{i: j} \equiv  k_i +k_{i+1} +\ldots+k_{j-1} +k_j,$.  For expressions involving only two (not necessarily consecutive) momenta, we are using the shorthand notation $P_{ij} \equiv k_i+k_j.$
We focus on the configuration where the sets of punctures $(\sigma_1,\sigma_2)$ and $(\sigma_3,\sigma_4)$ are respectably on the upper and the lower sheet of the curves \\[-15pt]
\begin{align}
(y_1=+\sqrt{\s_1^2-\L^2},\s_1), \quad (y_2=+\sqrt{\s_2^2-\L^2},\s_2),  \\[-3pt]
(y_3=-\sqrt{\s_3^2-\L^2},\s_3),  \quad (y_4=-\sqrt{\s_4^2-\L^2},\s_4). \nonumber
\end{align} \\[-10pt]
Expanding  all elements in $A_4'(1,2,3,4)$ around $\L=0$,  we obtain (to leading order)
\begin{align}
	{\rm PT}^\tau(1,2,3,4)\Big|^{1,2}_{3,4}= \frac{\L^2}{2^2} \frac{1}{(\s_{12} \s_{2P_{34}} \s_{P_{34}1} )} 
	\frac{1} { (\s_{P_{12}3} \s_{34} \s_{4P_{12}} )}, \nonumber
\end{align}
\begin{align}
&\frac{\Delta(123) \Delta(123|4)}{ S^{\tau}_4}
\Big|^{1,2}_{3, 4}  
= \\&\hskip0.8cm  \frac{2^5}{\L^4} (\s_{12}\, \s_{2P_{34}} \,  \s_{P_{34}1}  )^2 
 \left( \frac{ 1 }{ s_{34}} 
\right) \label{TTeq}
\times (\s_{P_{12}3} \, \s_{34} \, \s_{4P_{12}})^2,
 \nonumber \\
&
\prod_{a=1}^4   \frac{(y\s)_a}{y_a} \, T_{12}T_{23} \, {\rm det} [(\mathsf{A}^\L)^{12}_{23}]
\Big|^{1,2}_{3, 4}  \! =  \\ &\hskip2.9cm\frac{\L^2}{2^2} \frac{1} {\s_{12}\, \s_{2P_{34}}} \,\, 
\frac{s_{14} }{   \s_{P_{34}1} }
\frac{1}{\s_{P_{12}3}} \,\, \frac{(-1)\,s_{34}}{ \s_{34} \, \s_{4P_{12}} },\nonumber
\end{align}
where we have introduced the new fixed punctures $\s_{P_{34}}=\s_{P_{12}}=0$. Since we want to arrive at factorization identity for non-linear sigma model amplitudes (inspired by previous work for Yang-Mills theories, see ref. \cite{Gomez:2018cqg}) we are now going to introduce polarizations associated with the punctures, $\s_{P_{34}}=\s_{P_{12}}=0$,  {\it i.e.} $\epsilon^{M}_{34}$ and $\epsilon^{M}_{12}$. Thus,
\begin{align}
s_{14} & = 2(k_1\cdot k_4)=2\,(k_{1\,\mu} \times \eta^{\mu\nu} \times k_{4\,\nu})\\ &
=\sum_M (\sqrt{2}\,k_{1}\cdot \epsilon_{34}^M) \times (\sqrt{2}\,k_{4}\cdot \epsilon_{12}^M),\nonumber
\end{align}
employing,
\begin{align}
	\label{eq:epsM}
	\sum_M\epsilon_i^{M\, \mu} \epsilon_j^{M\, \nu} = \eta^{\mu\nu}.
\end{align}
After  separating the labels $\{1,2\}$ and $\{3,4\}$, it is simple to  rearrange the \cref{TTeq} as a product of two reduced determinants, 
\begin{align}\nonumber
 \frac{1} {\s_{12}\, \s_{2P_{34}}} \,  \frac{ (\sqrt{2} \, k_{1}\cdot \epsilon_{34}^{M})  }{   \s_{P_{34}1} }   =
 \frac{1} {\s_{12}\, \s_{2P_{34}}} \,
  {\rm det} \left[ \frac{\sqrt{2} \, k_{1}\cdot \epsilon_{34}^{M} }{   \s_{P_{34}1} } \right],  
\end{align}
\begin{align}\nonumber
\frac{(-1)}{\s_{P_{12}3}} \,\, \frac{s_{34} \, (\sqrt{2} \, k_{4}\cdot \epsilon_{12}^{M} ) }{ \s_{34} \, \s_{4P_{12}} }= 
 \frac{(-1)} {\s_{P_{12}3}} \,
  {\rm det} \left[ 
\begin{matrix}
\frac{\sqrt{2} \, k_{3}\cdot \epsilon_{12}^{M} }{   \s_{3P_{12}}} & \frac{s_{34}}{\s_{34}} \\
\frac{\sqrt{2} \, k_{4}\cdot \epsilon_{12}^{M} }{   \s_{4P_{12}}} & 0\\
\end{matrix}  
\right], 
\end{align}
therefore
\begin{eqnarray}
&&
\hspace{-0.1cm}
\prod_{a=1}^4   \frac{(y\s)_a}{y_a} \,\, T_{12}T_{23} \, {\rm det} [(\mathsf{A}^\L)^{12}_{23}]
\Big|^{1,2}_{3, 4}  =-
\frac{\L^2}{2^2}  \times   \sum_{M}      \frac{1} {\s_{12}\, \s_{2P_{34}}} \times
\nonumber \\ 
&& 
  {\rm det} \left[ \frac{\sqrt{2} \, k_{1}\cdot \epsilon_{34}^{M} }{   \s_{P_{34}1} } \right]\times
 \frac{(-1)} {\s_{P_{12}3}} \,
  {\rm det} \left[ 
\begin{matrix}
\frac{\sqrt{2} \, k_{3}\cdot \epsilon_{12}^{M} }{   \s_{3P_{12}}} & \frac{s_{34}}{\s_{34}} \\
\frac{\sqrt{2} \, k_{4}\cdot \epsilon_{12}^{M} }{   \s_{4P_{12}}} & 0\\
\end{matrix}  
\right].\label{neweqT}
\end{eqnarray}
The new matrices in eq. (\ref{neweqT}) can be obtained from the $\mathsf{A}$ matrix by replacing the off-shell momenta, $P_{34} $ and $P_{12}$, by their corresponding off-shell polarization vectors,
\begin{align}
& 
\!\!\!\!{\rm det} [ (\mathsf{A})^{12}_{2P_{34}} ]\to {\rm det} \left[ \frac{\sqrt{2} \, k_{1}\cdot \epsilon_{34}^{M} }{   \s_{P_{34}1} } \right] \ {\rm for} \ P_{34} \to \frac{1}{\sqrt{2} } \epsilon^{M}_{34}, \label{Nmatrix1}\!\!\!
\end{align}
\begin{align}
&
\!\!\!\!\! {\rm det}
[ (\mathsf{A})^{{P_{12}}}_{3} ] \to   {\rm det} \!\!\left[ 
\!\begin{matrix}
	\frac{\sqrt{2} \, k_{3}\cdot \epsilon_{12}^{M} }{   \s_{3P_{12}}} & \frac{s_{34}}{\s_{34}} \\
	\frac{\sqrt{2} \, k_{4}\cdot \epsilon_{12}^{M} }{   \s_{4P_{12}}} & 0\\
\end{matrix}  \!
\right]  {\rm for} \ {P_{12} \rightarrow \frac{1}{\sqrt{2} } \epsilon^{M}_{12} },
\label{Nmatrix2}
\end{align}
\noindent 
where the $\mathsf{A}$ matrix in \cref{Nmatrix1} is the $3\times 3$ matrix related with the punctures $(\s_1,\s_2,\s_{P_{34}})$, while the matrix in \cref{Nmatrix2} corresponds to the punctures  $(\s_{P_{12}},\s_3,\s_4)$. \\[5pt]
Using the measure, $d\mu_4^\L = \frac{1}{2^2}\frac{d\L}{\L}$, we now perform the $\L$ integral and the amplitude becomes
\begin{align}
& 
A_4'(1,2,3,4) \Big|^{1,2}_{3,4} =  \\
&
\frac{1}{2} \sum_M  \frac{A_3'(1,2,P^{\epsilon^M}_{34})
\times A_3^{(P_{12}3)}(P^{\epsilon^M}_{12},3,4)  }{s_{12}}  = \frac{s_{14}}{2} ,
\quad\nonumber
\end{align}
\noindent 
where the notation, $P_i^{\epsilon^M}$, means one must make the replacement, $P_i\rightarrow \frac{1}{\sqrt{2}}\, \epsilon^{M}_i$, 
and use \cref{eq:epsM}.
The overall factor $1/2$ cancels out after summing over mirrored configurations, {\it i.e.}, 
$A_4'(1,2,3,4) \Big|^{1,2}_{3,4} +A_4'(1,2,3,4) \Big|_{1,2}^{3,4} = s_{14}.$ \\
Following the integration rules in ref. \cite{Gomez:2018cqg}, we also have the contribution (up to summing over mirrored configurations) 
\begin{align}
& 
A_4'(1,2,3,4) \Big|^{4,1}_{2,3} =  \\
&
\frac{1}{2} \sum_M  \frac{A_3^{(1P_{23})}(1,P^{\epsilon^M}_{23},4) \times  
A_3'(P^{\epsilon^M}_{41},2,3)}{s_{14}}  = \frac{s_{12}}{2} .
\quad \nonumber
\end{align}
\noindent 
Thus, the final result is 
\begin{align}\label{4ptsAmp}
& 
A_4'(1,2,3,4) =  \\
&
\sum_M \left[  \frac{A_3'(1,2,P^{\epsilon^M}_{34})
	\times A_3^{(P_{12}3)}(P^{\epsilon^M}_{12},3,4)  }{s_{12}} \right. \nonumber \\
&
\left.  + \, \frac{A_3^{(1P_{23})}(1,P^{\epsilon^M}_{23},4) \times A_3'(P^{\epsilon^M}_{41},2,3)}{s_{14}} \right] = - s_{13}.
\qquad \nonumber
\end{align}
The four-point amplitude is factorized in terms of three-point functions. The general three-point
functions where some or all particles can be off-shell, are
\begin{align}
	A_3'(P_a, P_b, P_c) &= s_{P_c P_a} = -(P_a^2-P_b^2+P_c^2), \\
	A_3^{(P_a\,P_b)}(P_a,P_b,P_c) &= s_{P_b P_c}s_{P_c P_a} \\
	&=(P_c^2-P_a^2+P_b^2)(P_a^2-P_b^2+P_c^2). \nonumber
\end{align}
Since the non-linear sigma model is a scalar theory it is an interesting proposition to consider longitudinal degrees of freedom only
\begin{align}
	\label{eq:epsL}
\sum_{L}\epsilon_i^{L\,\mu}\epsilon_j^{L\,\nu}=\frac{k^\mu_ik^\nu_j}{k_i \cdot k_j}. 
\end{align}
Doing so we arrive at the equation
\begin{align}\label{4ptsAmpL}
& 
A_4'(1,2,3,4) =  \\
&
2\sum_L \left[ (-1)^3 \, \frac{A_3'(1,2,P^{\epsilon^L}_{34})
	\times A_3^{(P_{12}3)}(P^{\epsilon^L}_{12},3,4)  }{s_{12}} \right. \nonumber \\
&
\left.  +(-1)^3\,\frac{ A_3'(P^{\epsilon^L}_{41},2,3) \times A_3^{(1P_{23})}(1,P^{\epsilon^L}_{23},4)    }{s_{14}} \right] = -s_{13}.
\nonumber
\end{align}
\noindent 
Surprisingly, it is possible to generalize this equation to higher point amplitudes. Here the overall sign of each contribution depends of the number of points of the sub-amplitudes. In ref. \cite{preparation}, we will give more details on this phenomenon.  \\[5pt]
\section{New relations}
As will be shown in great detail elsewhere \cite{preparation}, 
using the double-cover prescription for a partial non-linear sigma model amplitude one is led to the following general formula where an $n$-point amplitude is factorized into a product of two (single-cover) lower-point amplitudes:
\begin{align}\label{GenFact}
& 
A_n' ( 1,2, 3,4,\ldots , n ) =     \nonumber\\
& 
\sum_{i=4,\, M}^{n}   \frac{A'_{n-i+3}\big(1,2,{P}^{\epsilon^{\!M}}_{\!3:i},i\!+\!1,... n\big)\, A^{(P_{i+1:2}3)}_{i-1}\big({{{P}^{\epsilon^{\!M}}_{\!i+1:2 }}},{3},... i \big) }{P_{i+1:2}^2}  \nonumber \\
& 
+\! \sum_{M} \frac{A'_{3}\big({{P}^{\epsilon^{\!M}}_{\!4:1}},2,3\big) \!\times \!A^{(1P_{23})}_{n-1}\big(1,{{P}^{\epsilon^{\!M}}_{\!23}},4,\ldots, n\big) }{ P_{23}^2} .
\end{align}
\noindent 
Here $n$ is an even integer and we have used \cref{eq:epsM}.
The above expression is valid using the M\"obius and scale-invariance gauge choice $(ijk|r)=(123|4)$.\\[5pt]
From the decomposition obtained by the double-cover method in \cref{GenFact}, we are able to write down a new factorization relation, where only longitudinal degrees of freedom contribute,
\\[-20pt]
\begin{align}\label{GenL}
& 
A_n' ( 1,2, 3,4,\ldots , n ) = 2\left[ \sum_{i=4 ,\, L}^{n}  (-1)^{i-1} \times  \right. \\
& 
\frac{A'_{n-i+3}\big(1,2,{P}^{\epsilon^{\!L}}_{\!3:i},i\!+\!1,... n\big)\times A^{(P_{i+1:2}3)}_{i-1}\big({{{P}^{\epsilon^{\!L}}_{\!i+1:2 }}},{3},... i \big) }{P_{i+1:2}^2} + \nonumber \\
& 
\left.
\sum_{L} (-1)^3 \frac{A'_{3}\big({{P}^{\epsilon^{\!L}}_{\!4:1}},2,3\big) \times A^{(1P_{23})}_{n-1}\big(1,{{P}^{\epsilon^{\!L}}_{\!23}},4,\ldots, n\big) }{ P_{23}^2} \right], \nonumber
\end{align}\\[-10pt]
\noindent 
where \cref{eq:epsL} was used. We checked this formula up to ten points.\\[5pt]
Since the above factorization relation includes only longitudinal contributions, we can rewrite it in a more elegant form, 
involving only the $A'_q$ amplitudes.
Using the definitions given in \cref{newAmp1,newAmp2} and  under the gauge fixing $(ijk)$, with $i<j<k$,  we have the following two identities \cite{preparation}\\[-15pt]
\begin{align}
A_q^{(ij)}(...,P_{i},...)
&=
P_i^2\,
A_q'(...,P_{i},...), ~ ~ q=2m+1 \nonumber
~~~ 
\\
A_q^{(ij)}(...,P_{i},...)
&=-
P_i^2\,
A_q'(...,P_{i},...), ~~  ~ q=2m, 
\label{iden1}
\end{align}\\[-15pt]
\noindent 
where $P_i^2 \neq 0$.
In addition, $A_q^{(ij)}$ satisfies the useful identities
\begin{align}
&
A_q^{(ij)}(1,...,i,...P_j,...k...q)
=
A_q^{(jk)}(1,...,i,...P_j,...k...q),\nonumber \\
&
A_q^{(ij)}(1...,i,...P_j,...k...q)
=A_q^{(ij)}(2...i...P_j...k...q,1) =\nonumber \\
&
\cdots =A_q^{(jk)}(...P_j...k...q...i) 
=A_q^{(ij)}(...k...q,1,...i...P_j...).
\label{iden2}
\end{align}
Applying the identities \cref{iden1,iden2}, it is straightforward to obtain
\begin{align}\label{GenL2}
& 
A_n' ( 1,2, 3,4,\ldots , n ) =   \\& \, \sum_{i=4 }^{n}      \frac{A'_{n-i+3}\big(1,2,{P}_{\!3:i},i\!+\!1,... n\big)\times\!  A'_{i-1}\big({{{P}_{\!i+1:2 }}},{3},... i \big) }{P_{i+1:2}^2}\nonumber  \\
& 
\hskip2cm+\,\frac{A'_{3} \big( P_{4:1},2,3\big) \times A'_{n-1}\big(P_{23},4,\ldots, n,1\big) }{ P_{23}^2},\nonumber 
\end{align}
\noindent 
where  the factorization formula has been written in terms of the generalized amplitude $A_q'$. Other gauge choices will naturally with lead to alternative factorization formulas.
\subsection{BCFW recursion}
It is interesting to analyse the new factorization identities in comparison with expressions originating from Britto-Cachazo-Feng-Witten (BCFW) recursion \cite{Britto:2005fq}. We introduce the momentum deformation
\begin{align}
k^\mu_2(z) = k^\mu_2 + z \,q^\mu, \quad k^\mu_3(z) = k^\mu_3 - z \,q^\mu, \quad z\in\mathbb{C},
\end{align}
\noindent 
where $q^\mu$ satisfies $k_2\cdot q=k_3\cdot q=q\cdot q=0$. 
Deformed momenta are conserved and on-shell: $k_1+k_2(z)+k_3(z)+k_4+\cdots +k_n=0$ and $k^2_2(z)=k^2_3(z)=0$. We consider the general amplitude, $A_n(1,\ldots, n)$, where $n$ is an even integer. From \cref{GenL2} using Cauchy's theorem we have 
\hskip1cm\begin{align}\label{residues1}
&
A_n(1,2,...,n)=\\ &
-\! \sum_{i=3 }^{n/2}    {\rm Res}_{ P_{2i:2}^2(z)=0} \nonumber
\Big[ 
A'_{n-2i+4}\big(1,2,{P}_{\!3:2i-1},2i,..., n\big)\times\!\!
\\
&\!\!
\left.   \frac{ A'_{2i-2}\big({{{P}_{2i:2 }}},{3},..., 2i-1 \big) }{z\,P_{2i:2}^2(z)}     
\right] \!-\!{\rm Res}_{z=\infty} \nonumber
\left[
\frac{A_n' (1,2,...,n) (z) }{z}
\right].
   \nonumber
\end{align}
\noindent 
Only the even amplitudes, namely $A_{2q}'$,  contribute to the physical residues. This is simple to understand as we have the identity,  
$A_{2q}(1,\ldots , 2q)=A_{2q}'(1,\ldots , 2q)$, so only sub-amplitudes with an even number of particles produce physical factorization channels. On the other hand, when the number of particles is odd, the off-shell ($P_i^2\neq0$) amplitude, $A_{2q+1}'(...,P_i,...)$, is proportional to $P_i^2$, since it must vanish when all particles are on-shell. So, the poles, $P^2_{2i-1:2}$, $i=3,...,\frac{n}{2}+1$ and $P_{23}$, are all spurious and the sub-amplitudes with an odd number of particles only contribute the boundary term at $z=\infty$.  \\[5pt]
Finally, it is important to remark  that after evaluating the residues, $P^2_{2i:2}(z)=0$, in \cref{residues1}, one obtains extra non-physical contributions, which cancel out combining with terms associated with the residue at $z=\infty$. Therefore, the effective boundary contribution is just given by the sub-amplitudes with an odd number of particles
\begin{align}\label{residueI}
&
 {\rm Res}_{z=\infty} 
 \left[
 \frac{A_n' (1,2,...,n) (z) }{z}
 \right]^{\rm Effective} \!\!\!=
 \\
&
 \partial_{\frac{1}{z}}
\left[ \sum_{i=3 }^{n/2+1}    
A'_{n-2i+5}\big(1,2,{P}_{\!3:2i-2},2i-1,..., n\big) \right.
   \nonumber\\
&
\left.
\times\,   \frac{ A'_{2i-3}\big({{{P}_{2i-1:2 }}},{3},..., 2i-2 \big) }{z\, P_{2i-1:2}^2(z)} \,+    
\right. \nonumber \\
&
\left.
\frac{A'_{3} \big( P_{4:1},2,3\big) \times A'_{n-1}\big(P_{23},4,\ldots, n,1\big) }{z\,  P_{23}^2} \right]_{z=\infty}.
\nonumber
\end{align}
\noindent 
\section{Conclusions} 
We have proposed a new CHY integrand for the $U(N)$ non-linear sigma model. For this new integrand, the kinematic matrix, $(\mathsf{A})^{ij}_{jk}$, is no longer anti-symmetric. We have found two new factorization identities, \cref{GenFact} and \cref{GenL}. We have written the second factorization formula in an elegant way, which only involves the generalized amplitude, $A'_q$. This formula turns out to be surprisingly compact (we have checked agreement of the soft-limit of this formula with ref.\cite{Cachazo:2016njl}).
\\[5pt]
This has implications for BCFW recursion since the two new factorization formulas can be split among even and odd sub-amplitudes, for example $A'_{2q} \times A'_{2m}$ and $A'_{2q+1} \times A'_{2m+1}$ respectively. Using this we are able to give a physical meaning to the odd sub-amplitudes as boundary contributions under such recursions.  \\[5pt]
Work in progress \cite{preparation} is  going to present a new recurrence relation and investigate  its connection to Berends-Giele \cite{Mizera:2018jbh,Berends:1987me,Chen:2013fya,Low:2017mlh,Kampf:2013vha} currents and Bern-Carrasco-Johansson (BCJ) numerators  \cite{Bern:2008qj,Du:2018khm,Carrasco:2016ldy}. 
Similar relations for others effective field theories \cite{Cachazo:2014xea,Mizera:2018jbh,Cachazo:2016njl} are expected and will be another focus. \\[5pt]
Despite similarities between the three-point amplitudes with the Feynman vertices obtained in 
ref. \cite{Cheung:2016prv}, the construction presented here is different. For example, the numerators found in \cref{4ptsAmp} are not reproduced by the Feynman rules found in ref. \cite{Cheung:2016prv}. Understanding the relationship between the formalisms would be interesting. \\[-3pt]
\begin{acknowledgements}
{\sc Acknowledgements:} ~Numerous discussions with J. Bourjaily and P. H. Damgaard are gratefully acknowledged. We thank C. Vergu for pointing out a useful identity.
This work was supported in part by the Danish National Research Foundation (DNRF91) and H. G. in part by the University Santiago de Cali  (USC).
\end{acknowledgements}

\end{document}